\documentclass[preprint,12pt]{elsarticle}

\usepackage{amssymb}
\usepackage{amsmath}
\usepackage{amsthm}
\usepackage{physics}

\usepackage{mathtools}
\usepackage{subfig}
\usepackage{booktabs} 
\usepackage{multirow}
\usepackage{graphicx, color}
\usepackage{xcolor, colortbl} 
\usepackage{adjustbox}

\usepackage[protrusion=true, expansion=true]{microtype}
\usepackage{subcaption}
\usepackage[normalem]{ulem}

\theoremstyle{plain}
\newtheorem{theorem}{Theorem}
\newtheorem{corollary}[theorem]{Corollary}

\theoremstyle{definition}

\usepackage{xr}

\newcommand\revision[1]{{\color{black}{{#1}}}}

\newcommand\minorrevision[1]{{\color{black}{{#1}}}}

\usepackage{hyperref}
\hypersetup{pdfauthor={Name}}
\journal{Computational Statistics \& Data Analysis}

\begin{document}

\begin{frontmatter}



\title{\textbf{Metric Oja depth, new statistical tool for estimating the most central objects}}


\author[label1]{Vida Zamanifarizhandi\corref{cor1}} 
\ead{vizama@utu.fi}

\author[label1]{Joni Virta} 


\cortext[cor1]{The corresponding author}

\affiliation[label1]{organization={Department of Mathematics and Statistics,\\ University of Turku},
            country={Finland}}

\begin{abstract}
The Oja depth (simplicial volume depth) is one of the classical statistical techniques for measuring the central tendency of data in multivariate space. Despite the widespread emergence of object data like images, texts, matrices or graphs, a well-developed and suitable version of Oja depth for object data is lacking. To address this shortcoming, a novel measure of statistical depth, the metric Oja depth applicable to any object data, is proposed. Two competing strategies are used for optimizing metric depth functions, i.e., for finding the deepest objects with respect to them. The performance of the metric Oja depth is compared with three other depth functions (half-space, lens, and spatial) in diverse data scenarios.
\end{abstract}




\end{frontmatter}



\section{Introduction}
\label{sec1}

\subsection{Background}
\label{subsec1}


    Nowadays, data are being generated in high volumes, at high speeds, and with great diversity. A key aspect of modern data analysis is that we no longer deal exclusively with data existing in Euclidean spaces. Instead, data are currently taking on more complex formats, such as images, graphs, matrices, etc., all of which reside in non-Euclidean spaces and are collectively known as \textit{object data}. Methodology for object data, known as object data analysis or metric statistics \citep{dubey2024metric}, are currently abundant in statistical literature; see, e.g., \cite{dubey2022modeling, virta2022sliced, zhang2023dimension, zhu2023geodesic}. Object data are typically modelled as residing in an arbitrary metric space $(\mathcal{X}, d)$ and we also adopt this approach.
    

     
     As with any data, the analysis of object data begins with exploratory data analysis on the raw data. This can take various forms: calculating descriptive statistics, dimension reduction, detecting outliers etc. However, despite being straightforward for Euclidean data, these tasks can be surprisingly difficult in the context of object data, due to the lack of structure in $(\mathcal{X}, d)$. In this work, we focus on one of the most fundamental exploratory statistical tasks, location/mean/average estimation, using \textit{depth functions} as our tool of choice.



    The foundation for depth functions was introduced in 1975 by Tukey who proposed his seminal half-space depth for multivariate data to rank observations and reveal features of the underlying data distribution~\citep{tukey1975mathematic}. Apart from this, numerous depth functions for data in Euclidean space have been proposed. E.g., the convex hull peeling depth (onion depth) \citep{barnett1976ordering, eddy1981graphics}, simplicial volume depth (Oja depth) \citep{oja1983descriptive} and several others; see Table \ref{tab1} and the comprehensive addressing in~\cite{mosler2022choosing}.

\begin{table}[h]
    \centering
    \resizebox{\textwidth}{!}{ 
    \begin{tabular}{lllcc}
      \hline
      \textbf{Author} & \textbf{Year} & \textbf{Depth function} & \textbf{Has a metric version} & \textbf{Complexity} \\
      \hline
      Mahalanobis \citep{mahalanobis1936generalized} &  1936 & Mahalanobis depth (distance)  & No & - \\
      Tukey \citep{tukey1975mathematic} & 1975 & Half-space/location/Tukey depth & Yes \citep{dai2023tukey} & $\mathcal{O}(n^2 h) + \mathcal{O}(n^3)$ \\
      Barnett, Eddy \citep{barnett1976ordering, eddy1981graphics} & 1976, 1981 & Convex hull peeling/onion depth & No & - \\
      Liu; Zuo \& Serfling \citep{liu1992data, zuo2000general} & 1992, 2000 & Projection depth & No & - \\
      Liu \citep{Liu1990} & 1990 & Simplicial depth & No & - \\
      Oja, Zuo \& Serfling \citep{oja1983descriptive} & 1983, 2000 & Simplicial volume depth/Oja depth & Yes, in the current paper & $\mathcal{O}(n^2 h) + \mathcal{O}(n^4)$ \\
      Koshevoy \& Mosler \citep{koshevoy1997zonoid} & 1997 & Zonoid depth & No & - \\
      Vardi \& Zhang, Serfling \citep{vardi2000multivariate, serfling2002depth} & 2000, 2002 & Spatial depth & Yes \citep{virta2023spatial} & $\mathcal{O}(n^2 h) + \mathcal{O}(n^3)$ \\
      Liu \& Modarres \citep{LiuModarres2011} & 2011 & Lens depth & Yes \citep{kleindessner2017lens, cholaquidis2023weighted, geenens2023statistical} & $\mathcal{O}(n^2 h) + \mathcal{O}(n^3)$ \\
      Yang \& Modarres \citep{yang2018beta} & 2018 & $\beta$-skeleton depths & \revision{Not for general $\beta$, see the caption} & - \\
      \hline
    \end{tabular}
    }
    \caption{\minorrevision{List of depth functions. In the bottom row, the metric $\beta$-skeleton depth is considered only for $\beta = 2$, where the concept reduces to lens depth. The listed complexities are for computing depths on the full training sample of size $n$; For metric half-space depth, the complexity assumes the use of \cite[Algorithm~1]{dai2023tukey} with the sample itself as anchors. The complexities consist of two parts, distance matrix formation and depth computation, where in the former, $h$ denotes the cost of computing the distance between a single pair of objects in $(\mathcal{X}, d)$; see Section \ref{subsec:sample_level} for further details.}}
    \label{tab1}
\end{table}


In recent years, object/metric versions of various depth functions $D$ have been proposed; see the second-to-last column of Table \ref{tab1}. That is, given an object $X \in \mathcal{X}$ and a distribution $P$ taking values in $\mathcal{X}$, the depth $D(X; P)$ describes how central the object $X$ is w.r.t. $P$, much in the same way as for the Euclidean depths earlier. Two key properties of these extensions are that (a) they depend on the data only through the metric $d$, making them applicable to \textit{any} form of object data, regardless in which specific metric space they live, and (b) if $(\mathcal{X}, d)$ is a Euclidean space, then the original Euclidean version of the depth is recovered, showing that these metric depth functions are ``true'' generalizations of their classical counterparts. Metric lens depth was developed in \cite{kleindessner2017lens, cholaquidis2023weighted, geenens2023statistical}, metric half-space depth was proposed in \cite{dai2023tukey} and metric spatial depth in \cite{virta2023spatial}.

Most of the depth functions in Table \ref{tab1}, in particular all three that have object versions, are \textit{robust} \citep{maronna2019robust}, meaning that their performance is not skewed by the presence of possible outliers. \revision{In Section \ref{sec:review} we list the robustness properties of the existing metric depth functions, and more details on the robustness of Euclidean depths can be found in \cite[Table 1]{mozharovskyi2025anomaly} and \cite{zuo2000general, mozharovskyi:tel-03780415}}. Note that not all location estimation procedures are robust. For example, the Fr{\'e}chet mean \citep{frechet1948elements}, which is a generalization of the concept of average and a classical estimator of location for object data, is not robust, but is instead affected greatly by data outliers. Despite the popularity of Fr{\'e}chet mean and similar tools, in this work we concentrate solely on robust methods. This is because object data are very diverse, which makes recognizing any possible outliers very difficult. Robust analysis methods protect against the influence of outliers and leads to reliable results, regardless of the exact type of outliers and whether they are detected in the data or not.

\subsection{Our Contributions}

This work focuses on the robust location estimation of object data using depth functions, and its main contributions are as follows.


(i) We propose a new depth function for object data, the metric Oja depth, and extensively study its theoretical properties. In particular, we characterize its range of values and show the sample consistency of its maximizers under specific assumptions.



(ii) We study the estimation of the deepest objects (data location). That is, given a sample $X_1, \ldots, X_n$ of object data, we aim to find the object $X \in \mathcal{X}$ which has the maximal depth value w.r.t. the sample. Despite the fundamental nature of this problem, in the earlier works on metric depths it has been considered only by \cite{dai2023tukey}, \revision{who proposed finding the optimum among the sample points $X_1, \ldots, X_n$, or using more elaborate manifold optimization to find the deepest out-of-sample object, as implemented in the R-package \texttt{MHD} \cite{RMHD}}. As a competing approach, we propose using non-linear Euclidean optimizers along with a PCA-reduced coordinate representation of the object data to estimate the deepest out-of-sample object. 

(iii) We conduct extensive simulation and data comparisons between our proposed depth and its competitors in estimating the deepest object in several data scenarios. In particular, we investigate how much the in-sample estimation can be improved by using non-linear optimization to achieve out-of-sample estimation. The results show that, at the cost of greater computational complexity, metric Oja depth surpasses its competitors in several scenarios, especially when combined with our proposed optimization strategies.



\subsection{Contents}



We review the three existing metric depth functions in Section~\ref{sec:review}. In Section \ref{sec:new_depth}, we introduce the proposed new metric depth function and develop its theoretical properties. In Section \ref{sec:simu}, we evaluate and compare it with three other depth functions, using two simulation scenarios. In Section \ref{sec:real_data}, we further test our metric depth function on a real dataset, using permutation and rank tests to obtain inferential results. Finally, future recommendations are compiled in Section \ref{sec:discussion}.

\section{Review of existing metric depth functions}\label{sec:review}


Let $(\Omega, \mathcal{F}, \mathbb{P})$ be a probability space and take $(\mathcal{X}, d)$ to be a complete and separable metric space where our data reside. \revision{Recall that a metric \( d: \mathcal{X} \times \mathcal{X} \to \mathbb{R} \) quantifies distances between objects residing in the set $\mathcal{X}$. Formally, metrics are required to satisfy the following axioms: Positivity, $d(x, y) \geq 0$ with equality if and only if $x = y$; Symmetry, $d(x, y) = d(y, x)$; and Triangle inequality, $d(x, y) \leq d(x, z) + d(z, y)$ \citep{burago2001metric, gromov2007metric}.} Further, we let $P$ be a probability measure defined on the Borel sets of $\mathcal{X}$. Given a fixed, non-random object $x \in \mathcal{X}$, the metric depth functions answer the question ``how central is the object $x$ with respect to the distribution $P$?''.

To tackle this question, \cite{kleindessner2017lens, cholaquidis2023weighted, geenens2023statistical} proposed metric lens depth, defined as
\begin{align*}
    D_L(x) := \mathbb{P} \left( d(X_1, X_2) > \max \left\{ d( X_1, x), d(X_2, x) \right\} \right),
\end{align*}
where $X_1, X_2$ are independent random variables with the distribution $P$. Drawing an analogy from Euclidean statistics, $D_L(x)$ essentially gives the probability that the side corresponding to $X_1, X_2$ is the longest in a ``triangle'' drawn using the objects $x, X_1, X_2$. Intuitively, the probability is small (large) if $x$ is located far away from (close to) the bulk of the distribution $P$, making $D_L$ behave as expected from a depth function.


\cite{dai2023tukey} gave a similar treatment to the classical half-space depth and defined the metric half-space depth as
\begin{align*}
    D_{H}(x)& = \inf_{{x_1,x_2} \in \mathcal{X}, \ d(x_1, x) \leq d(x_2, x)} \mathbb{P} (d(X, x_{1}) \leq d(X, x_{2})).
\end{align*}
Every pair of objects $x_1, x_2$ in the infimum divides the space $\mathcal{H}$ in two subsets (objects closer to $x_1$ than $x_2$ and vice versa). The depth $D_H(x)$ is defined to be the smallest possible probability mass of a subset produced in this way and containing the object $x$. As with the metric lens depth, it is easy to see that, for an outlying object $x$, it is easy to find a subset which contains $x$ but has very small $P$-probability mass, making $D_H(x)$ small. An approximative algorithm, that we also use in this paper to compute the metric half-space depth, is given in \cite[Algorithm 1]{dai2023tukey}.

The metric spatial depth was proposed in \cite{virta2023spatial} as

\begin{align*}
    \resizebox{\textwidth}{!}{$D_S(x) = 1 - \frac{1}{2} \mathbb E \left[ \mathbb{I}(d(X_1, x) \neq 0, d(X_2, x) \neq 0) \left\{ \frac{ d^2(X_1, x) + d^2(X_2, x) - d^2(X_1, X_2) }{d(X_1, x) d(X_2, x)} \right\} \right],$}
\end{align*}
where $X_1, X_2 \sim P$ are independent. The interpretation of $D_S(x)$ is trickier than for the previous two depths, but it essentially measures how likely $x$ and two random objects $X_1, X_2$ are to yield equality in the triangle inequality. The depth takes values in $[0, 2]$ and the value $0$ is reached if and only if the probability of equality is 1 and $x$ is never in the middle of the other two objects, whereas to reach the value $2$, $x$ must always reside between the two other objects. More details on interpreting the metric spatial are given in~\cite{virta2023spatial}.


\revision{Throughout this work, we refer to metric lens depth, metric half-space depth and metric spatial depth as MLD, MHD and MSD, respectively.} All of these have the property of \textit{vanishing at infinity} meaning that, for objects far enough away from the bulk of the distribution $P$, the depth values approach zero. All three also satisfy specific forms of continuity (small changes in $x$ cause small changes in the depth) and both MLD and MHD are invariant to data transformations that preserve the ordering of distances. Regarding robustness properties, MHD and MSD are robust, the former in the sense of having a high breakdown point and the latter in the sense of having a bounded influence function. Whereas, MLD is also likely to be highly robust, being based only on comparisons between distances and not their absolute sizes, but this aspect of MLD has not yet been studied in the literature. For detailed statements of these, and some other properties of the three depth functions, including sample consistency results, we refer the reader to the papers \cite{cholaquidis2023weighted, geenens2023statistical, dai2023tukey, virta2023spatial}.



\section{New depth for object data}\label{sec:new_depth}


\subsection{Population-level properties}

\revision{For any three objects $x_1, x_2, x_3 \in \mathcal{X}$, we use the notation $L(x_1, x_2, x_3)$ to denote the event that $d(x_1, x_3) = d(x_1, x_2) + d(x_2, x_3)$. On an intuitive level, $L(x_1, x_2, x_3)$ means that $x_2$ resides ``between'' $x_1$ and $x_3$, since under this event the distance from $x_1$ to $x_3$ is the same as when taking a detour through $x_2$. Hence, in the sequel we use phrases such as ``$x_2$ lies in between $x_1$ and $x_3$'' to indicate that $L(x_1, x_2, x_3)$ holds. For arbitrary objects $x_0, x_1, x_2, x_3 \in \mathcal{X}$, we denote by $B_3(x_0, x_1, x_2, x_3)$ the $3 \times 3$ matrix whose $(k, \ell)$-element equals
\begin{align}\label{eq:B3_element_definition}
   \frac{1}{2} \left\{ d^2(x_0, x_k) + d^2(x_0, x_\ell) - d^2(x_k, x_\ell) \right\}.
\end{align}
}





Letting $X_1, X_2, X_3 \sim P$ be i.i.d. random objects, we propose to measure the depth of a fixed object $x \in \mathcal{X}$ with respect to the distribution $P$ as
\small
\begin{align}\label{eq:do3}
    D_{O3}(x) := \frac{1}{ 1 + \mathbb E \left[ \left\{ | B_3(x, X_1, X_2, X_3) | + 4 d^2(x, X_1) d^2(x, X_2) d^2(x, X_3) \right\} ^{1/2} \right] }.
\end{align}
\normalsize
The subscript O refers to ``Oja'' due to the close connection between $D_{O3}(x)$ and the classical Oja depth explained in Appendix A. By Theorem B.1 in Appendix B, taking the square root in the denominator of \eqref{eq:do3} is well-defined. The intuitive meaning of $D_{O3}(x)$ is not clear from its definition \eqref{eq:do3} alone, so we next present a collection of results that allow better understanding it, starting with a moment condition that guarantees its existence.

\begin{theorem}\label{theo:moment_condition}
    Assume that, for some $a \in \mathcal{X}$, we have $\mathbb E \{ d(X, a) \} < \infty$. Then $D_{O3}(x)$ exists as well-defined.
\end{theorem}

By ``well-defined'' in Theorem \ref{theo:moment_condition} we mean that the expected value used to compute $D_{O3}(x)$ exists as finite. The condition $\mathbb E \{ d(X, a) \} < \infty$ is analogous to requiring a univariate random variable to have a finite mean and is a rather mild condition as even the Frech{\'e}t mean already requires the existence of second moments. The choice of $a$ in Theorem \ref{theo:moment_condition} is completely arbitrary as, if the condition $\mathbb E \{ d(X, a) \} < \infty$ holds for some $a$, then, by the triangle inequality, it holds for all $a \in \mathcal{X}$.

Theorems \ref{theo:L_interpretation} and \ref{theo:vanishing_at_infinity} next establish the range of $D_{O3}(x)$ to be $[0, 1]$ and give interpretations to these endpoints.

\begin{theorem}\label{theo:L_interpretation}
    $D_{O3}(x)$ takes values in $[0, 1]$. Moreover, if $D_{O3}(x) = 1$, then
    \begin{align*}
        P\{ L(X_1, x, X_2) \cup L(X_2, x, X_3) \cup L(X_3, x, X_1) \} = 1,
    \end{align*}
    where $X_1, X_2, X_3 \sim P$ are i.i.d.
\end{theorem}

Theorem \ref{theo:L_interpretation} implies that large values of $D_{O3}(x)$ are indicative of the centrality of the object $x$ with respect to the distribution $P$. That is, under the extremal case $D_{O3}(x) = 1$ the object $x$ must almost surely lie in between of at least two of three randomly sampled objects $X_1, X_2, X_3$ from $P$.

We say that a sequence of objects $x_n \in \mathcal{X}$ is divergent if there exists $a \in \mathcal{X}$ such that $d(x_n, a) \rightarrow \infty$ when $n \rightarrow \infty$. Intuitively, a divergent sequence is such that it moves ``towards infinity'' eventually getting arbitrarily far from any fixed object. Clearly, such sequences do not exist in every metric space.


\begin{theorem}\label{theo:vanishing_at_infinity}
Assume that, for some $a \in \mathcal{X}$, we have $\mathbb E \{ d^2(X, a) \} < \infty$. Let $x_n$ be a divergent sequence of objects in $\mathcal{X}$. Then, $D_{O3}(x_n) \rightarrow 0$ as $n \rightarrow \infty$.
\end{theorem}

The second moment condition in Theorem \ref{theo:vanishing_at_infinity} is used to avoid certain pathological behavior; see the proof of the result in Appendix B. Theorem \ref{theo:vanishing_at_infinity} states that small values of $D_{O3}(x)$ are reached by outlying objects, i.e., points that are located far away from the bulk of the distribution $P$.

It is typical for depth functions to satisfy various invariance properties \cite{dai2023tukey, geenens2023statistical}. Namely, one usually expects that the depth of a point $x$ w.r.t. to distribution $P$ does not change if $x$ and $P$ are both subjected to a specific class of transformations. Being based on pair-wise distances, $D_{O3}$ inherits such properties directly from the metric $d$. That is, if $d(g x_1, g x_2) = d(x_1, x_2)$ for all $g \in \mathcal{G}$ where $\mathcal{G}$ is a group of transformations on $\mathcal{X}$, then also $D_{O3}$ is invariant to $\mathcal{G}$. In particular, if $x_0 \in \mathcal{X}$ has maximal depth w.r.t. $P$, then $g x_0$ has maximal depth w.r.t. to $P_g$, the distribution of $g X$ where $X \sim P$.

\subsection{Sample-level properties}\label{subsec:sample_level}

Let $X_1, \ldots, X_n$ be a random sample of objects from $P$. To estimate depths in practice, a natural sample counterpart of $D_{O3}$ is the third-order $U$-statistic,
\begin{align}\label{eq:d03n_formula}
    D_{O3, n}(x) := \frac{1}{1 + \frac{1}{\binom{n}{3}} \sum_{(i, j, k)} h(x, X_i, X_j, X_k)},
\end{align}
where the summation runs over all unordered triples $(i, j, k)$ of indices from $\{ 1, \ldots, n \}$ with distinct elements and
\begin{align*}
    h(x, X_1, X_2, X_3) := \{ | B_3(x, X_1, X_2, X_3) | + 4 d^2(x, X_1) d^2(x, X_2) d^2(x, X_3) \}^{1/2}.
\end{align*}
Standard results on $U$-statistics \citep{lee2019u} combined with the continuous mapping theorem and Theorem \ref{theo:moment_condition} give the following point-wise convergence result which states that the depth of any single object $x$ can be accurately estimated using $D_{O3, n}$.

\begin{theorem}\label{theo:point_wise_convergence}
    Assume that, for some $a \in \mathcal{X}$, we have $\mathbb E \{ d(X, a) \} < \infty$. Then, for any fixed $x \in \mathcal{X}$, we have $D_{O3, n}(x) \rightarrow_p D_{O3}(x)$ as $n \rightarrow \infty$.
\end{theorem}

A stronger result is obtained if the metric space is totally bounded, in which case the convergence is uniform in $\mathcal{X}$.

\begin{theorem}\label{theo:uniform_convergence}
    Assume that the metric space $(\mathcal{X}, d)$ is totally bounded. Then,
    \begin{align*}
        \sup_{x \in \mathcal{X}} | D_{O3, n}(x) - D_{O3}(x) | \rightarrow_p 0,
    \end{align*}
    as $n \rightarrow \infty$.
\end{theorem}

The classical $M$-estimator argument \cite[Theorem 5.7]{van2000asymptotic} then shows that the point with maximal depth, assuming that it is unique, can be consistently estimated with the sample depth function $D_{O3, n}$.

\begin{corollary}\label{cor:deepest_convergence}
    Assume that the metric space $(\mathcal{X}, d)$ is totally bounded and that $x \mapsto D_{O3}(x)$ has unique maximizer $x_0$. Then any sequence $x_n \in \mathcal{X}$ of maximizers of $x \mapsto D_{O3, n}(x)$ satisfies $x_n \rightarrow_p x_0$ as $n \rightarrow \infty$.
\end{corollary}

The sample maximizers in Corollary \ref{cor:deepest_convergence} exist since $D_{O3, n}$ is a continuous function on a totally bounded and complete, and hence compact, set. The assumption of a unique population maximizer is discussed further in Section~\ref{sec:discussion}. Also MLD and MHD satisfy analogous consistency properties; see \cite{cholaquidis2023weighted, geenens2023statistical, dai2023tukey}.





\minorrevision{We conclude the subsection by discussing computational aspects of the proposed $D_{O3, n}$. Obtaining the depths of all sample points comprises of (A) computing the distance matrix $\{ d(X_i, X_j) \}_{i,j = 1}^n$ and, (B) using this to compute the depths of the sample points with \eqref{eq:d03n_formula}. The cost of step (A), distance computation, depends on the metric space in question and is essentially $O(n^2 h)$ where $h$ is the cost of computing the distance between a single pair of objects in $(\mathcal{X}, d)$. For example, both in $\mathbb{R}^p$ with the Euclidean metric and on the unit sphere in $\mathbb{R}^p$ with the arc length metric, forming the distance matrix is an $O(n^2 p)$-operation, whereas on the manifold of $p \times p$ positive definite matrices with either the log-Euclidean or Riemannian metric, this operation is $O(n^2 p^3)$. Importantly, step (A) is identical for all metric depth functions, implying that their complexities can be compared based on step (B) only, detailed next.} 

\minorrevision{Step (B), computing the actual depths, is done based on the $n \times n$ distance matrix and, as such, the complexity of this step depends only on $n$ and not on other aspects of the sample. In particular, the dimensionality of the data plays no role in this step. The computational complexity of our proposed depth is from \eqref{eq:d03n_formula} seen to be $\mathcal{O}(n^4)$, whereas the complexity of MLD, MSD and MHD is $\mathcal{O}(n^3)$, for computing the depths of the full sample (assuming \cite[Algorithm 1]{dai2023tukey} with the sample itself as anchors is used for MHD). The time efficiency of our proposed depth could be improved, at the cost of accuracy, by using partial U-statistics to randomly choose which elements to include in the triple sum in \eqref{eq:d03n_formula}. However, we have not resorted to this compromise here. The previous information on the complexities has been summarized also in the final column of Table \ref{tab1}. The complexities of the Euclidean counterparts of these depths and other Euclidean depths can be found in \cite[Table 1.9]{mozharovskyi:tel-03780415}.}
 
\minorrevision{The computation of MLD and MHD does not rely on the actual distances $d(X_i, X_j)$ but only on indicators of the form $\mathbb{I}\{ d(X_i, X_j) \leq d(X_i, X_k)\}$; see Section~\ref{sec:review}. This means that MLD and MHD (a) are invariant to transformations that preserve these indicators, and (b) can be computed even if one knows only the binary values of these triple comparisons. The latter viewpoint for MLD was taken in \cite{kleindessner2017lens} to find deepest objects based on binary dissimilarity data. Our proposed depth, on the other hand, uses the actual values $d(X_i, X_j)$ in its computation, meaning that it cannot be computed based on indicators alone. This limits its usefulness in such dissimilarity scenarios but, at the same time, allows our depth to use additional information from the data, which later manifests as improved performance in several experiments compared with the other methods. This behavior is perfectly in line with the original Oja depth, which also depends on the numerical values of the distances $\| X_i  - X_j \|$.}




\subsection{Auxiliary depth-like quantity}\label{sec:auxiliary}

\revision{
Our proposed depth $D_{O3}$ is based on the 3-dimensional Oja depth; see Appendix A for details, and a natural question is whether also the $p$-dimensional Oja depth can be used for similar purposes with $p \neq 3$. In this section we briefly investigate the version with $p = 2$. Letting $B_2(x_0, x_1, x_2)$ denote the $2 \times 2$ top left sub-matrix of $B_3(x_0, x_1, x_2, x_3)$, we define
\begin{align}\label{eq:depth_O2}
    D_{O2}(x) := \frac{1}{1 + \mathbb E \left\{ | B_2(x, X_1, X_2) |^{1/2} \right\} },
\end{align}
which, as shown in Appendix A, is a metric generalization of the classical 2-dimensional Oja depth. As with $D_{O3}$, also $D_{O2}$ is well-defined as soon as the first moment exists, i.e., $\mathbb E \{ d(X, a) \} < \infty$ for some $a \in \mathcal{X}$ (proof omitted). However, the next result, which follows directly from Theorem B.1(i) in Appendix B shows that $D_{O2}(x)$ does not actually measure the centrality of $x$ w.r.t. $P$ and is thus not a proper depth measure.

\begin{theorem}\label{theo:U_interpretation}
    $D_{O2}(x)$ takes values in $[0, 1]$. Moreover, $D_{O2}(x) = 1$ if and only if
    \begin{align*}
        P\{ L(x, X_1, X_2) \cup L(X_2, x, X_1) \cup L(X_1, X_2, x) \} = 1,
    \end{align*}
    where $X_1, X_2 \sim P$ are i.i.d.
\end{theorem}

Intuitively, $D_{O2}(x)$ takes a large value if, for random $X_1, X_2$ and the fixed object $x$, one of the three objects is typically located between the other two. This, however, does not guarantee that $x$ is central with respect to $P$, as it is possible that $X_1$ (or $X_2$) is the in-between point. Thus, large values of $D_{O2}(x)$ do not characterize centrality. As an extreme example, if $(\mathcal{X}, d)$ is the one-dimensional Euclidean space, then $D_{O2}(x) = 1$ for all $x \in \mathbb{R}$. Consequently, we do not pursue the theoretical properties of $D_{O2}$ further. However, we still include $D_{O2}(x)$ in our data examples, where it performs surprisingly well in locating the centers of distributions.

}


\section{Simulation examples}\label{sec:simu}

    \subsection{In-sample optimization}\label{subsec:insample}

    Throughout the paper, we use the following abbreviations for the depth functions: MOD3 (metric Oja depth), MOD2 (the depth-like counterpart of MOD3 defined in Section~\ref{sec:auxiliary}), MHD (metric half-space depth), MLD (metric lens depth), and MSD (metric spatial depth). The depths were implemented in \texttt{Rcpp} \citep{eddelbuettel2018extending} and are available at \url{https://github.com/vidazamani/Depth-functions-for-Object-Data}.

    In our first experiment, we compare the depths with simulations in the context of location estimation. Given a sample $X_1, \ldots, X_n \in \mathcal{X}$ from a distribution $P$, each sample depth function $D_n$ is used to determine an in-sample estimate of the deepest object, i.e., $\hat{\mu} := X_{i_0}$ where $i_0 = \mathrm{argmin}_{i = 1, \ldots, n} D_n(X_i)$. The depths are compared based on the averages, over 200 replications, of the estimation errors $d(\hat{\mu}, \mu)$ where $\mu \in \mathcal{X}$ is the true location of the distribution $P$. In addition to the error, we also evaluate the computation times of the depths. \minorrevision{When measuring the running times, we disregard the computation of the interobject distances, as this step is common to all competing methods.}

    \subsubsection{Correlation matrix dataset}\label{cormat}
    \sloppy

    In our first simulation, we consider the Riemannian manifold $(\mathcal{X}, d)$ of $p \times p$ correlation matrices as the data space, where $d(X_1, X_2) = \| \mathrm{Log}(X_1^{-1/2} X_2 X_1^{-1/2}) \|_F$ is the affine invariant Riemannian metric; see \cite{bhatia2009positive}. Such data are commonly encountered, e.g., in neuroscience in the form of brain connectivity matrices~\citep{simeon2022riemannian}.
    


    The random $p \times p$ correlation matrices $X_1, \ldots , X_n$ were generated as $X_i := \mathrm{diag}(S_i)^{-1/2} S_i \mathrm{diag}(S_i)^{-1/2}$, where $S_i := U_i D_i U_i'$, $U_i$ is a random orthogonal matrix and $D_i$ is a diagonal matrix with all positive diagonal elements, independent of $U_i$. The diagonal matrices $D_i$ were generated as follows
    \[
    D_i = \mathrm{diag}\left(\exp(\mathcal{N}(\nu,1)), \underbrace{\exp(\mathcal{N}(-\nu,1)), \ldots, \exp(\mathcal{N}(-\nu,1))}_{p-1 \text{ times}}\right).
    \]
    The parameter $\nu$ was determined as follows: each $X_i$ had the probability $\varepsilon > 0$ of being an outlier, in which case $\nu = 3$, whereas the remaining $1 - \varepsilon$ proportion of the data (the bulk) used $\nu = 0$. The matrices $U_i$ are uniformly random; see \cite{Stewart1980} for details. This choice ensures that the eigenvectors of $S_i$ do not favor any direction. \revision{Consequently, the distribution of $S_i$ is rotationally symmetric around the half-line $\{ C I_p \mid C \geq 0 \}$ and, intuitively, the most central matrix w.r.t. this distribution has to reside on this half-line. As the correlation matrices $X_i$ are obtained from the $S_i$ by scaling the diagonal elements to unity, any reasonable depth measure should estimate the central object of the distribution of $X_i$ to be $\mu = I_p$.} Note that both the bulk and the outliers share the same true most central object (identity matrix) but the high variance of the outliers makes the estimation more difficult for larger $\varepsilon$.


\begin{figure}
    \centering
    \includegraphics[width=1\linewidth]{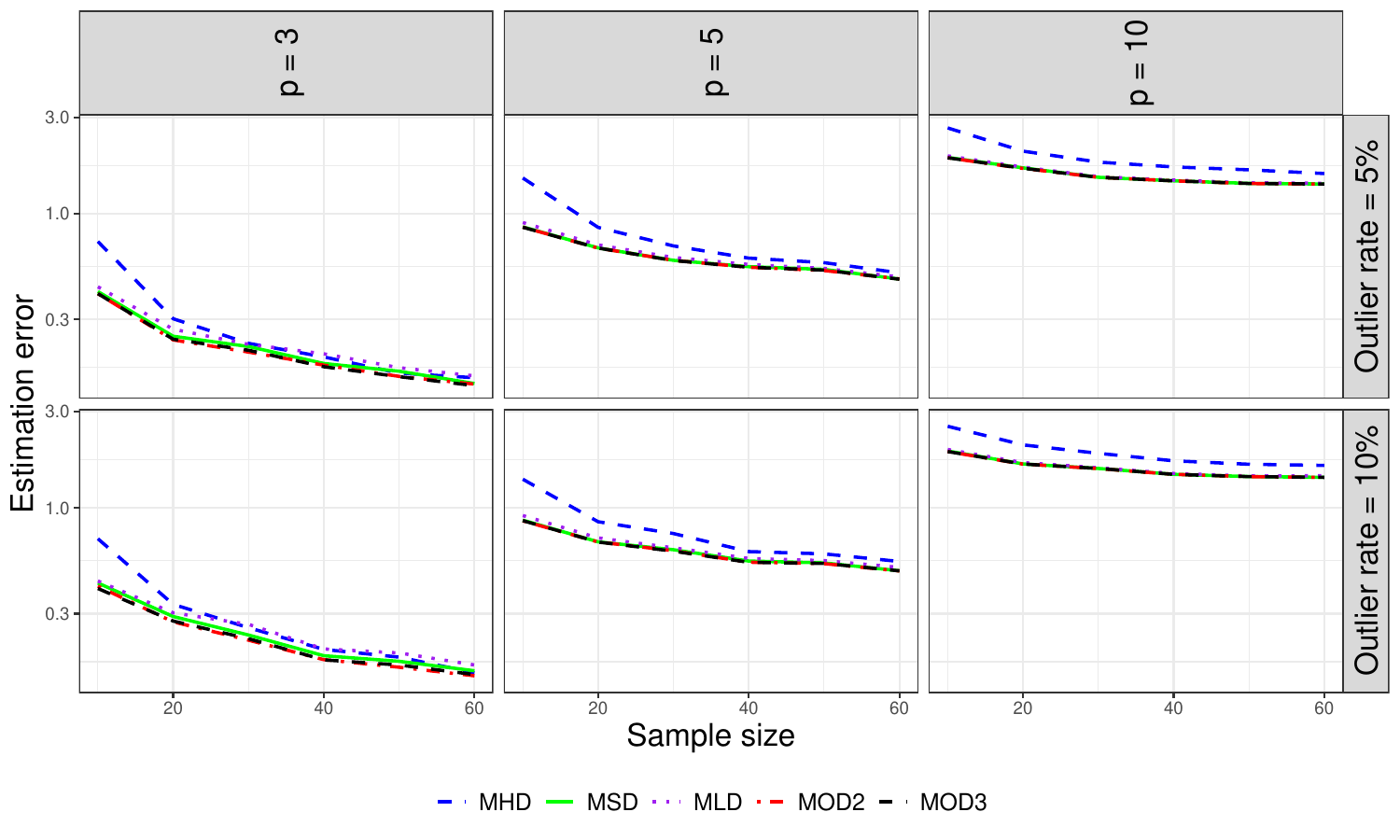}
    \caption{\revision{The average estimation errors for each of the five methods in the correlation matrix simulation. The scale of the $y$-axis is logarithmic.}}
    \label{fig:est_error_corr}
\end{figure}



\begin{figure}
        \centering
        \includegraphics[width=1\linewidth]{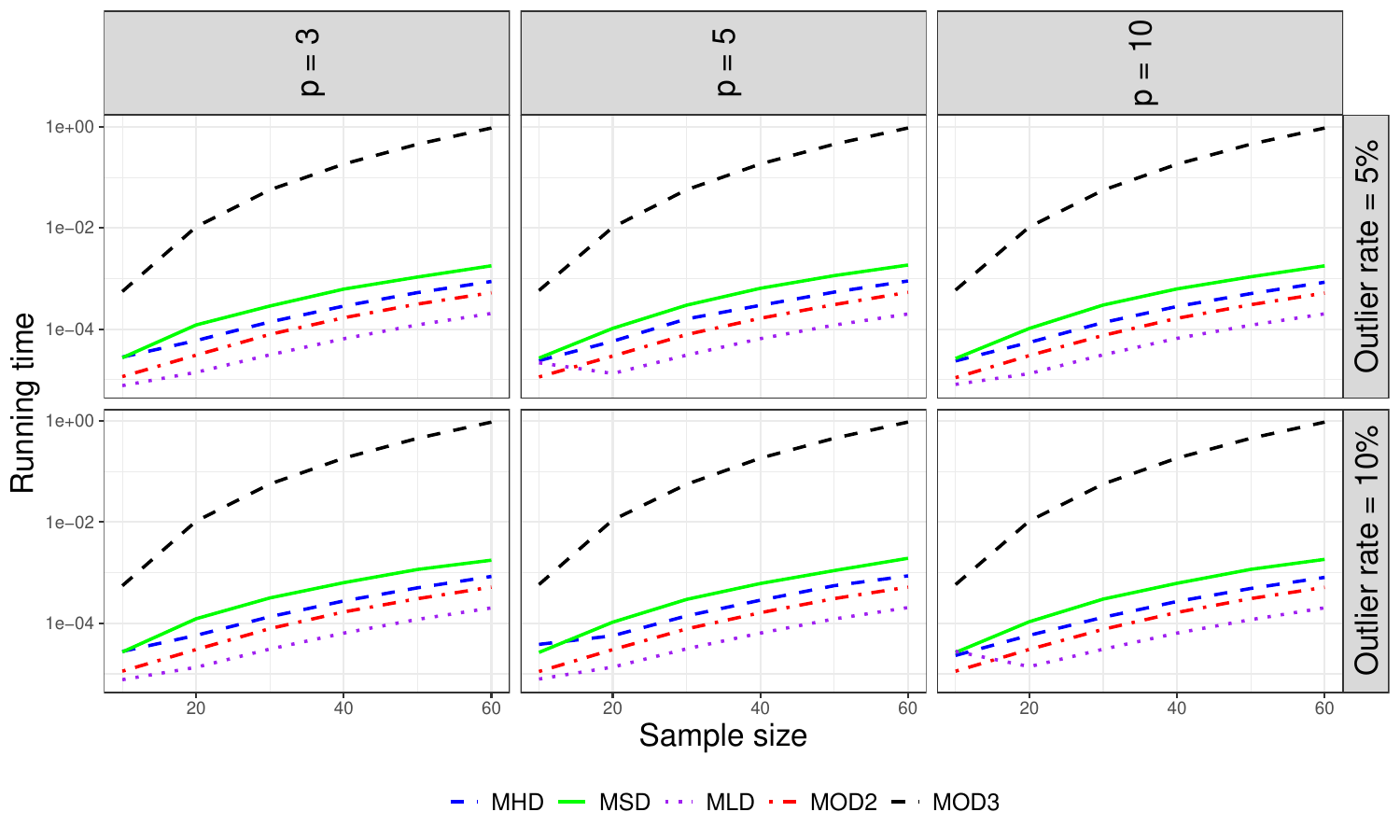}
        \caption{\revision{The running times (in seconds) of each method in the correlation matrix simulation. The scale of the $y$-axis is logarithmic.}}
        \label{fig:time_corr}
    \end{figure}

    
    The simulation has three parameters, the contamination proportion $\varepsilon = 0.05, 0.10$, dimension $p = 3, 5, 10$ and sample size $n = 10, 20, 30, 40, 50, 60$, leading to a total of 36 cases. Note that $n$ was deliberately chosen as low, since object data scenarios typically have much smaller sample sizes than in Euclidean data analysis. The average estimation errors are shown in Figure~\ref{fig:est_error_corr}, grouped by $\varepsilon$ and $p$. In all cases, as $n$ increases, the estimation error gradually decreases. Moreover, the complexity of the data space grows as the dimension $p$ of the matrices increases, leading to a decline in accuracy. Increasing the percentage of outliers from $5\%$ to $10\%$ did not lead to significant changes in the results, indicating the stability and robustness of the estimators. In all cases, MOD3 and MOD2  demonstrated the best performance, respectively, while MHD showed the weakest performance. MSD was the third most efficient method and close to MOD3 and MOD2 in performance. From the timing results given in Figure~\ref{fig:time_corr}, we observe that neither $p$ nor $\varepsilon$ affect the running time. This is because neither has an effect on the size of the distance matrix; see the discussion in Section \ref{subsec:sample_level}. Additionally, MOD3 and MOD2 have a relatively high time cost; see Figure \ref{fig:time_corr}, meaning that these depths essentially offer improved performance at the cost of speed. An alternative version of Figure \ref{fig:est_error_corr} where the results are shown relative to MSD is presented in Appendix C.








    \subsubsection{Hypersphere dataset}\label{hypersph}


    In this experiment, we generated samples of points on a $p$-dimensional unit hypersphere $\mathcal{X}$. As a metric $d$, we used the usual arc length distance. Each point $X_i$ was simulated by first generating the $p$-dimensional random vector $Z_i \sim \mathcal{N}_p(\lambda 1_p, I_p)$, where $1_p$ is a vector of ones and $\lambda \neq 0$, and then taking $X_i := Z_i/\| Z_i \|$. By symmetry, this strategy leads to the most central object always being equal to $\mu = \mathrm{sign}(\lambda) (1/\sqrt{p}) 1_p$ and the absolute value of the parameter $\lambda$ controls the spread of the points: the larger the value of $| \lambda |$, the more closely the points are concentrated around $\mu$. For the outliers we used $\lambda = -1$ and for the bulk $\lambda = 5$, meaning that the two distributions have their most central points on the opposite sides of the hypersphere; see Figure C.1 in Appendix C for an illustration.



    \begin{figure}
        \centering
        \includegraphics[width=1\linewidth]{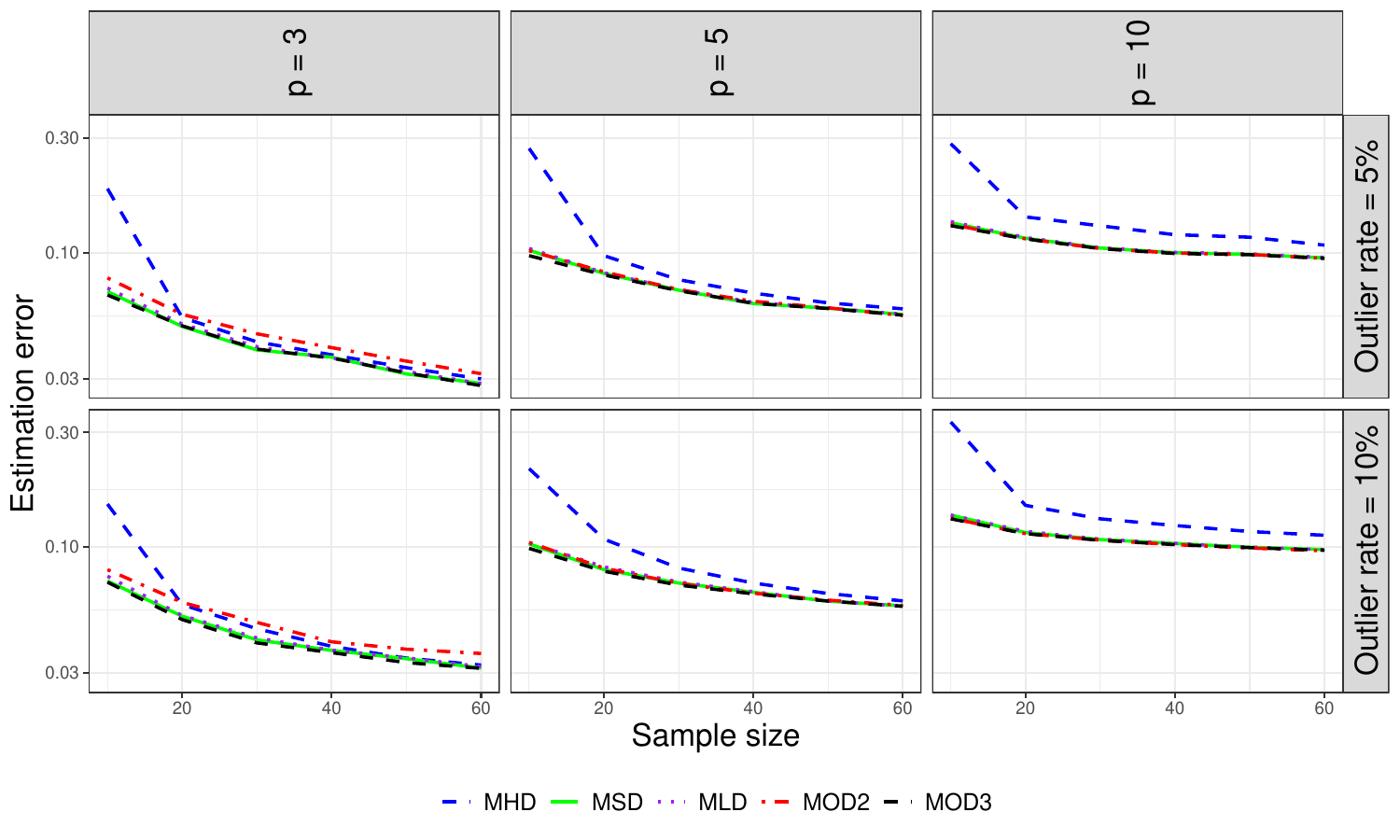}
        \caption{\revision{The average estimation errors for each of the five methods in the hypersphere simulation. The scale of the $y$-axis is logarithmic.}}
        \label{fig:est_error_sph}
    \end{figure}


    The simulation thus has three parameters, $\varepsilon$, $p$ and $n$, which are set to the same values as in the earlier simulation. As shown in Figure \ref{fig:est_error_sph} and its relative version presented in Appendix C, the performance of each metric depth function on the hypersphere dataset is very similar to its performance on the correlation data. We have omitted the timing results for the sphere simulation, as they were visually almost identical to those in Figure \ref{fig:time_corr}.
    
\begin{figure}
    \centering
    \includegraphics[width=0.6\linewidth]{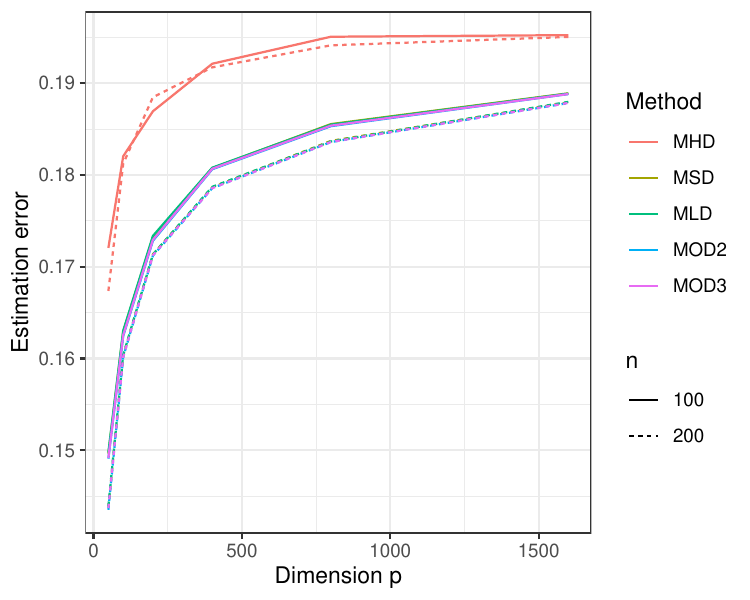}
    \caption{The effect of increasing the dimension of the unit sphere to the in-sample estimation error, estimated using 100 replications. The lines of MSD, MLD, MOD2 and MOD3 are almost perfectly overlapping.}
    \label{fig:p_experiment}
\end{figure}

    \revision{We also conducted an additional simulation in the same setting with $\varepsilon = 0.10$, $n = 100, 200$, and $p = 50, 100, 200, 400, 800, 1600$ to investigate the effect of increasing the data dimensionality $p$ on the estimate. The average estimation errors over 100 replicates are plotted in Figure~\ref{fig:p_experiment}. The results indicate that: (i) MSD, MLD, MOD2, MOD3 yield almost identical results and, as expected, their performance improves as $n$ increases. (ii) MHD is worse than the other four and changing $n$ does not appear to affect it. The standard half-space depth is known to degenerate in higher dimensions \citep{dutta2011some} and it is therefore possible that MHD suffers from the same issue. A non-trivial solution could be to regularize its computation in a suitable way. (iii) For all methods, the estimation error increases with $p$. From the viewpoint of efficiency, for example $n = 100$ and $p = 500$ yields roughly the same accuracy as $n = 200$ and $p = 650$. (iv) As a baseline, we also considered an estimator which picks a random object from the sample and uses it as the estimate of the location. Regardless of $p$, this leads to estimation errors of $0.413$ and $0.4078$ for $n = 100$ and $n = 200$, respectively, showing that all estimators improve considerably upon this simple benchmark.}

\subsection{Out-of-sample optimization}\label{outofsam}

    The in-sample optimization studied in the previous section might come across as a naive approach, but in many metric spaces it is the best one can do, the lack of Euclidean structure preventing the use of standard optimization algorithms. However, despite being non-Euclidean, some metric spaces admit coordinate representations/encodings in Euclidean spaces. Such spaces include, e.g., unit spheres (stereographic projection), the positive definite manifold (Cholesky decomposition) and the space of $L_2$-functions (Karhunen-Lo{\`e}ve expansion, assuming we truncate it). In these cases, Euclidean non-linear optimization can be used to find the deepest point.
    
    To formalize the earlier, assume that there exists a bijective map $f: \mathcal{X} \to \mathcal{S}$ where $\mathcal{S} \subseteq \mathbb{R}^q$ for some $q$. Then, the problem of finding the deepest object of a sample $X_1, \ldots, X_n \in \mathcal{X}$ can be formulated as
    \begin{align*}
        \max_{v \in \mathcal{S}} D_n( f^{-1}(v) ),
    \end{align*}
    where $D_n$ is some sample metric depth function and $v$ is an Euclidean vector. After finding the optimizer $v_0$, it can be mapped to $\mathcal{X}$ simply as $f(v_0)$. Since the map $v \mapsto D_n( f^{-1}(v) )$ is highly non-linear, \revision{we propose using bounded \textit{Nelder-Mead} (NMKB) and \textit{Limited-memory BFGS} (L-BFGS-B) for the optimization. NMKB is a robust and derivative-free optimizer, which makes it generally well-suited for non-smooth functions like metric depths. However, it can be slow in high-dimensional problems. On the other hand, L-BFGS-B uses approximated gradients to find the optimum solutions and stores only the most recent steps during the optimization process, making it faster and more memory-efficient. See \cite{BFGS1995} for more details.




    We employed here the implementation provided by the R-package \texttt{dfoptim} for NMKB and R’s base \texttt{optim()} function for L-BFGS-B, configuring the algorithm with the following tuning parameters. The initial values were set using the top five in-sample deepest points and the search space was defined with a width of 0.10 for each dimension. To further alleviate computational burden, we apply a dimension reduction via PCA to the representations in the space $\mathcal{S}$ to map them to an $r$-dimensional space with $r \ll q$. As training data for the dimension reduction, we use the images $f(X_1), \ldots ,f(X_n)$ of the sample.  As data, we generate a sample of correlation matrices similarly as described in Section \ref{cormat}, meaning that $(\mathcal{X}, d)$ is now the manifold of positive definite matrices having unit diagonal. We define the map $f$ such that $f(X_i)$ is the vector of length $p(p + 1)/2$ containing the non-zero elements in the Cholesky decomposition of $X_i$. Since the used optimization algorithms can produce solutions lying outside of the set $\mathcal{S}$ (that is, solutions which are covariance matrices, but not correlation matrices), we manually scale the final optimizer $f(v_0)$ into a proper correlation matrix.

    In this simulation, the matrix dimension $p$ was set to 5, and the contamination proportion $\varepsilon$ was fixed to 10\%. The sample size, as in previous simulations in Sections \ref{cormat} and \ref{hypersph}, was taken to be $n = 10, 20, 30, 40, 50, 60$. A new parameter introduced in this simulation was the number of principal components, determined by a threshold $ \texttt{TSH} \in (0, 1] $, representing the minimum proportion of total variance to be explained by the chosen principal components. Thus, the number of principal components is $r = \max ( 2, \{ k : \sum_{i=1}^k \text{ExplainedVariance}_i \geq \text{TSH} \} )$. In this simulation \texttt{TSH} was set to be $0.9$. The whole simulation was repeated 500 times using both MHD and MOD3. As further competitors we took the MHD-based in-sample and out-of-sample estimators in the R-package \texttt{MHD} \cite{RMHD}; see also \cite{dai2023tukey}. The latter is based on Riemannian optimization of the depth function and denoted by DLP hereafter. Finally, also the in-sample version of MOD3 was included.

\begin{figure}
    \centering
    \includegraphics[width=0.92\linewidth]{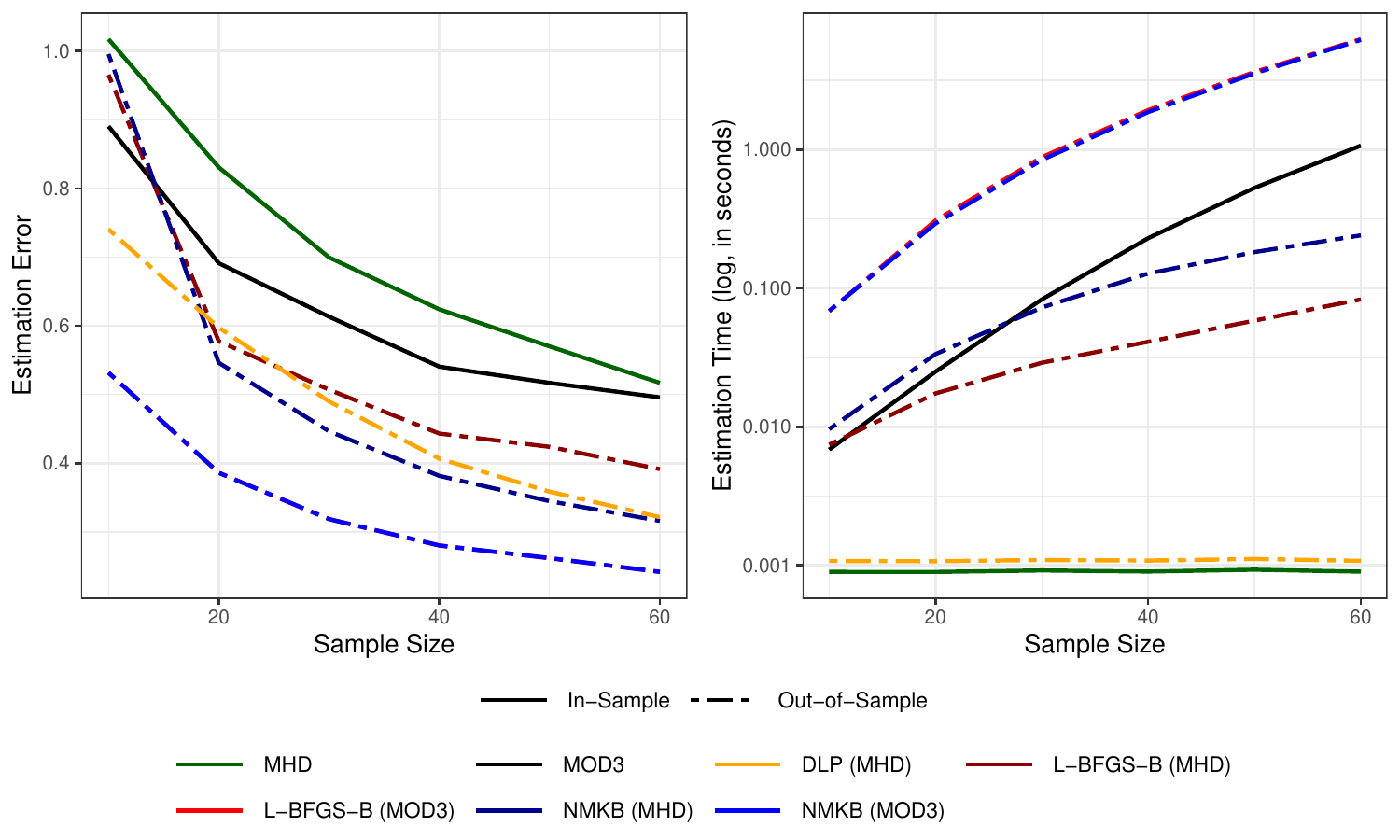}
    \caption{\revision{Comparing the performance of different optimization methods on MOD3 and MHD. DLP refers to the out-of sample method. Left: estimation error of 7 different cases. L-FBSG-B yielded exactly the same result on MOD3 as NMKB (light red line). Right: Estimation Time. The scale of the y-axis is logarithmic.}}
    \label{modmhdcomp}
\end{figure}

    As observed in the results in Figure \ref{modmhdcomp} (left), the use of the two introduced optimization algorithms (NMKB and L-BFGS-B) combined with dimension reduction has significantly contributed to reducing the estimation error, compared to using the in-sample estimators indicated by the solid lines. We further observe that NMKB has the best performance, surpassing the other two optimization strategies, DLP and L-BFGS-B. Moreover, MOD3 exhibits lower estimation error than MHD in both out-of-sample and in-sample cases. However, as shown in the Figure~\ref{modmhdcomp} (right), MOD3 involves a higher computational cost compared to MHD. Additionally, despite the substantial computational burden imposed by using the NMKB algorithm, we also observe that the MOD3 in-sample estimator tends to follow a similar trend to its optimization counterparts. 



}



\section{Real data example}\label{sec:real_data}



In this section, we evaluate the performances of the proposed depth functions using real data. The secondary aim of this experiment is to demonstrate how the depth functions, whose analytically complex form prevents theoretical inferential results, can still be used for statistical inference via relying on permutation and rank tests. \minorrevision{Similar depth-based tests have also been considered earlier in the context of complex and high-dimensional data. For instance, \cite{LopezPintadoWrobel2017} proposed several robust two-sample permutation tests for imaging data based on multivariate volume depth. Moreover, a depth-based global two-sample envelope test was developed in \cite{LopezPintadoQian2021} for functional and image data, and \cite{dai2023tukey} used depth-based rank testing proposed in \cite{ChenouriSmall2012} to compare samples of functional connectivity matrices.}

For this experiment, the \texttt{Age\_pyramids\_2014} data from R-package \texttt{HistDAWass} were chosen. This dataset contains histogram-valued data representing the age distribution of the population in each country worldwide, reported for both sexes combined as well as separately for males and females. We model the histograms as objects in the Wasserstein space $(\mathcal{X}, d)$ where $d$ is the $L_2$-Wasserstein distance, implemented in the \texttt{HistDAWass}-package.

\begin{figure}[t!]
    \centering
    \includegraphics[width=0.9\linewidth]{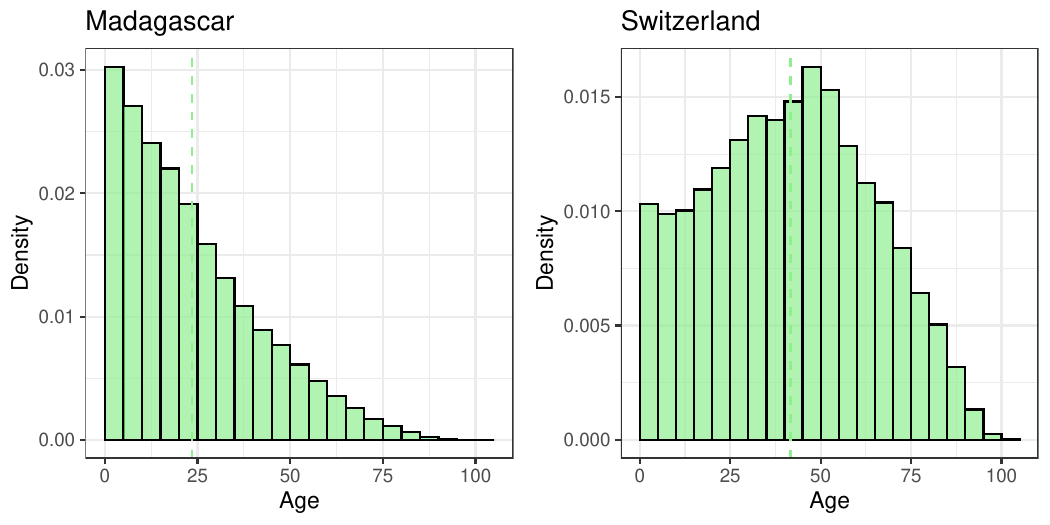}
    \caption{The deepest age histograms among African (left) and European (right) countries, with respect to $L_2$-Wasserstein distance and MOD3. The dashed green lines depict the means of the two distributions.}
    \label{barrevorg}
\end{figure}

\begin{figure}[t]
    \centering
    \includegraphics[width=1\linewidth]{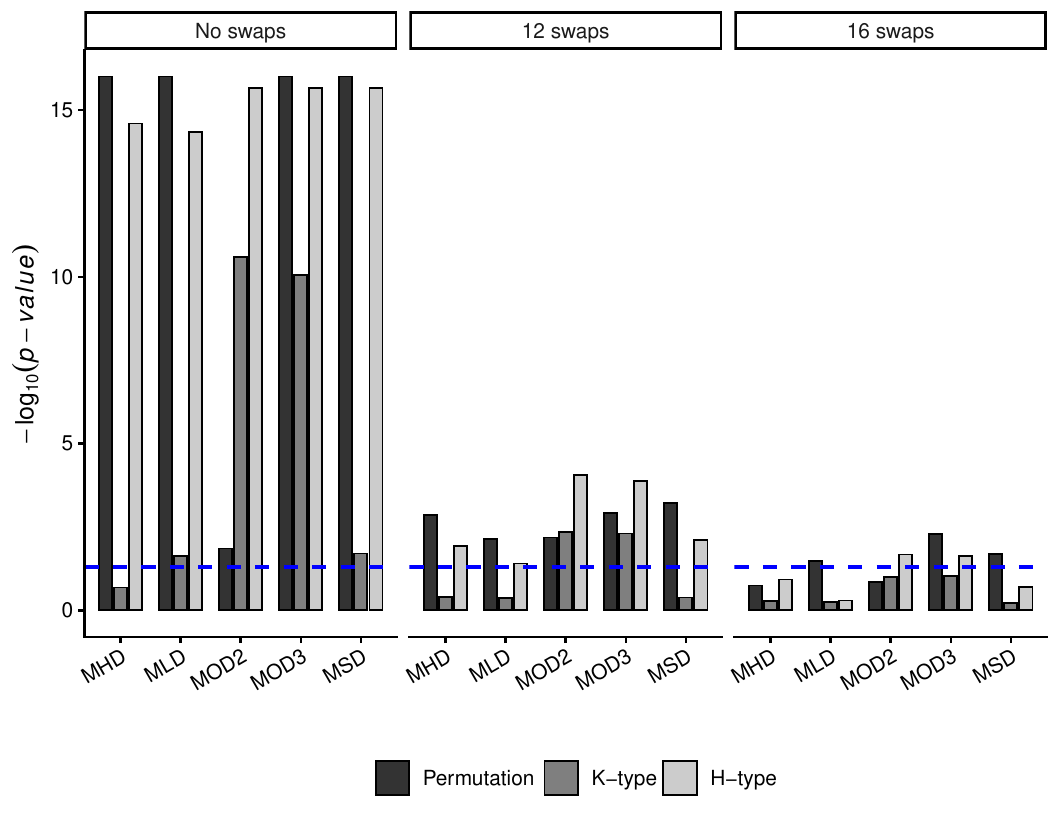}
    \caption{$p$-values of depth-based permutation and rank tests across all metric depth functions in the case of no contamination (left panel), 12 swaps (middle panel) and 16 swaps (right panel). Significance level of $-\log_{10}(0.05)$ is indicated by blue dashed line. All $p$-values smaller than $10^{-16}$ were recorded as $10^{-16}$.}
    \label{rankpermfast}
\end{figure}

The current analysis was conducted on a subset of data that includes both sexes across all African and European countries, accounting for a total of 95 countries. Based on the geographic locations of the two continents, we hypothesize that their age distributions differ.
  This is visually confirmed in Figure \ref{barrevorg}, where the deepest age distribution, w.r.t. MOD3, among African countries (left) shows a different structure from the corresponding country in Europe (right). Thus, our research question is: Is there a significant difference in the age distribution between these two groups? To study this, we next conduct a permutation test to assess whether the difference between the deepest histograms in the two continents is statistically significant. Letting $f_1, f_2 \in \mathcal{X}$ denote the deepest in-sample histograms of the two groups, we use the $L_2$-Wasserstein distance between them, $t = d(f_1, f_2)$, as the test statistic. 
  To simulate the distribution of $t$ under the null hypothesis (i.e., assuming no difference between the two groups), we randomly permute the group labels of the $n = 95$ histograms and compute the test statistic value. This reshuffling was done a total of 500 times, yielding $t^*_1, \ldots, t^*_{500}$, using which the $p$-value of the test is computed as $\#\{ t \leq t^*_i \}/500$. \minorrevision{Because the experiment produces very small $p$-values, we rescaled the $y$-axis in Figure~\ref{rankpermfast} to display $-\log_{10}(p\text{-value})$ instead. Under this transformation, taller bars correspond to smaller $p$-values and therefore provide stronger evidence against the null hypothesis. All depth functions indicate a difference between the groups at the $0.05$ significance level; see the black bars in Figure~\ref{rankpermfast} (left).}

\revision{We next take the previous result of a difference between the continents as a ground truth and continue the experiment by (a) contaminating the original data set, and (b) performing the same test for the contaminated data set and observing which of the depths still let us obtain the correct conclusion (group difference). This experiment therefore mimics the practical scenario where the data have been partially wrongly recorded. To perform the contamination, we randomly select $k$ countries and swap their labels (a European country is labeled an African country
and vice versa), using two different choices, $k = 12, 16$. To reduce randomness in the choice of the swap, we repeat the simulation 10 times, using 500 iterations in the test each time and the resulting average $p$-values are shown as black bars in Figure \ref{rankpermfast} (middle and right). Under medium contamination (12 out of 95 countries mislabeled), all depths continue to yield the correct decision. Whereas, under high contamination (16 out of 95 countries mislabeled) both MHD and MOD2 identify the groups as indistinct. This is likely due to the fact that the algorithm used for MHD is approximate \citep{dai2023tukey}. In addition, MOD2 is not a ``true'' depth function and therefore does not necessarily measure the centrality of its input object; see the discussion after Theorem \ref{theo:U_interpretation}. The most consistent performance across swaps is given by MOD3 and MSD, which likely relates to the fact that they both use numerical distance information in their computation, unlike MHD and MLD which only use indicators; see Section \ref{subsec:sample_level}. The results also align with the simulation studies in Section \ref{subsec:insample} where MSD was almost always as good as MOD3.}




\minorrevision{Moreover, following \cite{ChenouriSmall2012, dai2023tukey}, we implemented depth-based rank tests using MLD, MHD, MSD, MOD2, and MOD3 for the same null hypothesis of no difference between the continents. The idea behind the tests is to inspect the ranks of the observation depths, which should show no group pattern under the null hypothesis. Two versions of the test described in~\cite{ChenouriSmall2012} were considered: the K-statistic and the H-statistic. The K-statistic, essentially a Wilcoxon rank-sum test, uses pooled depth rankings but may lose power under location-shift alternatives (in the Euclidean case). The H-statistic is instead a more powerful alternative in which the depth rankings are computed separately relative to each group and then averaged across groups. Being depth-based, both statistics inherit all invariance properties of the underlying depth function. For both statistics, we used the asymptotic null distributions described in \cite{ChenouriSmall2012}.



\newpage
The $p$-values of the rank tests are shown in dark grey (K-statistic) and light grey (H-statistic) in Figure \ref{rankpermfast}. As seen in the left panel, the depth-based rank test rejects the null hypothesis for the original data in all models except one: MHD under the K-type statistic does not lead to rejection at the conventional significance level. The corresponding numerical values are reported in Table~C.1 in the appendix. This indicates that the permutation test provides more uniform sensitivity across depths, whereas the rank-based approach is more affected by the choices between the depths and the K- and H-type statistics. The results of the rank test under the swapping experiment in Figure \ref{rankpermfast} (middle and right) show that only MOD2 and MOD3, when combined with the H-type statistic, manage to reject the null under the strongest contamination (16 swaps). This is in line with the claim in \cite{ChenouriSmall2012}, which reports that the K-type test is the less powerful of the two rank tests. Overall, among the depth functions considered, MOD3 appears to be the most stable, showing strong performance under both permutation and rank-based testing frameworks.

}


\section{Conclusions}\label{sec:discussion}

    \revision{Possible future research topics include: (a) Selecting and setting appropriate values for the hyper-parameters (take initial values as an example) used in the optimization algorithms in Section \ref{outofsam}. (b) Investigating why $D_{O2}$, despite not being a measure of depth, manages to emprically estimate central objects on the manifold of positive definite matrices and on the unit sphere. (c) Depth functions are typically expected to be maximized at a symmetric center of a distribution. However, unlike, e.g., the metric half-space depth, $D_{O3}$ does not have a natural companion concept of symmetry using which such a center could be defined. One possibility would be to use the quantities $ A(x, X_1, X_2) := \{d^2(x, X_1) + d^2(x, X_2) - d^2(X_1, X_2)\}/\{2 d(x, X_1) d(x, X_2)\} $, which satisfy $A(x, X_1, X_2) \geq -1$ with equality if and only if $x$ lies in between of $X_1$ and $X_2$ \cite[Theorem 1]{virta2023measure}. This suggests that if there exists an object $x_0$ such that
\begin{align}\label{eq:stoch_dom}
    A(x_0, X_1, X_2) <_S A(x, X_1, X_2), \quad \mbox{for all } x \in \mathcal{X}\setminus \{ x_0 \}, 
\end{align}
where $<_S$ denotes a suitable form of strict stochastic dominance, then $x_0$ is a natural candidate for the most central point. Informally, \eqref{eq:stoch_dom} means that $x_0$ is more likely to satisfy $d(X_1, X_2) \approx d(X_1, x_0) + d(x_0, X_2)$ than any other point in $\mathcal{X}$. Accordingly, we conjecture that if $x_0$ satisfies \eqref{eq:stoch_dom}, then it also uniquely maximizes $D_{O3}$.}

\subsection*{Acknowledgments}

The work of VZ was supported by the Finnish Doctoral Program Network in Artificial Intelligence, AI-DOC (decision number VN/3137/2024-OKM-6). The work of VZ and JV was supported by the Research Council of Finland (Grants 347501, 353769, 368494).

\subsection*{Supplementary materials}

The supplementary appendices contain proofs of the technical results, additional simulation plots, and other material. The methods and simulation code (in R) are accessible at \url{https://github.com/vidazamani/Depth-functions-for-Object-Data}. 

\bibliographystyle{elsarticle-harv} 
\bibliography{referencesR1}

@article{LopezPintadoWrobel2017,
  author  = {L{\'o}pez-Pintado, Sara and Wrobel, Julia},
  title   = {Robust nonparametric tests for imaging data based on data depth},
  journal = {Stat},
  volume  = {6},
  number  = {1},
  pages   = {405--419},
  year    = {2017}
}

@article{LopezPintadoQian2021,
  author  = {L{\'o}pez-Pintado, Sara and Qian, Ke},
  title   = {A depth-based global envelope test for comparing two groups of functions with applications to biomedical data},
  journal = {Statistics in Medicine},
  volume  = {40},
  number  = {7},
  pages   = {1639--1652},
  year    = {2021}
}

@article{ChenouriSmall2012,
  author  = {Chenouri, Soumaya and Small, Christopher G.},
  title   = {A Nonparametric Multivariate Multisample Test Based on Data Depth},
  journal = {Electronic Journal of Statistics},
  year    = {2012},
  volume  = {6},
  pages   = {760--782}
}

@article{mosler2022choosing,
  title={Choosing Among Notions of Multivariate Depth Statistics},
  author={Mosler, Karl and Mozharovskyi, Pavlo},
  journal={Statistical Science},
  volume={37},
  number={3},
  pages={348--368},
  year={2022},
  publisher={Institute of Mathematical Statistics}
}

@article{zuo2000general,
  author    = {Zuo, Yijun and Serfling, Robert},
  title     = {General Notions of Statistical Depth Function},
  journal   = {Annals of Statistics},
  pages     = {461-482},
  year      = {2000},
  volume    = {28},
  number    = {2}
}

@article{barnett1976ordering,
  author    = {Barnett, V.},
  title     = {The Ordering of Multivariate Data},
  journal   = {Journal of the Royal Statistical Society, Series A},
  volume    = {139},
  number    = {3},
  pages     = {318--344},
  year      = {1976},
  note      = {With discussion}
}

@article{eddy1981graphics,
  author    = {Eddy, W. F.},
  title     = {Graphics for the Multivariate Two-Sample Problem: Comment},
  journal   = {Journal of the American Statistical Association},
  volume    = {76},
  number    = {374},
  pages     = {287--289},
  year      = {1981},
}

@article{frechet1948elements,
  author = {Fréchet, M.},
  title = {Les éléments aléatoires de nature quelconque dans un espace distancié},
  journal = {Annales de l'Institut Henri Poincaré},
  volume = {10},
  year = {1948},
  pages = {215--310}
}

@article{zhu2023geodesic,
  title={Geodesic Optimal Transport Regression},
  author={Zhu, Changbo and M{\"u}ller, Hans-Georg},
  journal={Accepted for publication in Biometrika},
  year={2025}
}

@article{dubey2022modeling,
  title={Modeling Time-Varying Random Objects and Dynamic Networks},
  author={Dubey, Paromita and M{\"u}ller, Hans-Georg},
  journal={Journal of the American Statistical Association},
  volume={117},
  number={540},
  pages={2252--2267},
  year={2022},
  publisher={Taylor \& Francis}
}

@article{zhang2023dimension,
  title={Dimension Reduction for {F}r{\'e}chet Regression},
  author={Zhang, Qi and Xue, Lingzhou and Li, Bing},
  journal={Journal of the American Statistical Association},
  pages={1--15},
  volume={119},
  year={2024},
  publisher={Taylor \& Francis}
}

@article{virta2022sliced,
  title={Sliced Inverse Regression in Metric Spaces},
  author={Virta, Joni and Lee, Kuang-Yao and Li, Lexin},
  journal={Statistica Sinica},
  volume={32},
  pages={2315--2337},
  year={2022}
}

@article{oja1983descriptive,
  title={Descriptive Statistics for Multivariate Distributions},
  author={Oja, Hannu},
  journal={Statistics \& Probability Letters},
  volume={1},
  number={6},
  pages={327--332},
  year={1983},
  publisher={Elsevier}
}

@article{geenens2023statistical,
  title={Statistical Depth in Abstract Metric Spaces},
  author={Geenens, Gery and Nieto-Reyes, Alicia and Francisci, Giacomo},
  journal={Statistics and Computing},
  volume={33},
  number={2},
  pages={46},
  year={2023},
  publisher={Springer}
}

@article{cholaquidis2023weighted,
  title={Weighted Lens Depth: Some Applications to Supervised Classification},
  author={Cholaquidis, Alejandro and Fraiman, Ricardo and Gamboa, Fabrice and Moreno, Leonardo},
  journal={Canadian Journal of Statistics},
  volume={51},
  number={2},
  pages={652--673},
  year={2023},
  publisher={Wiley Online Library}
}

@article{dai2023tukey,
  title={Tukey’s Depth for Object Data},
  author={Dai, Xiongtao and Lopez-Pintado, Sara},
  journal={Journal of the American Statistical Association},
  volume={118},
  number={543},
  pages={1760--1772},
  year={2023},
  publisher={Taylor \& Francis}
}

@book{maronna2019robust,
  title={Robust Statistics: Theory and Methods (with {R})},
  author={Maronna, Ricardo A and Martin, R Douglas and Yohai, Victor J and Salibi{\'a}n-Barrera, Mat{\'\i}as},
  year={2019},
  publisher={John Wiley \& Sons}
}

@article{virta2023measure,
  title={Measure of shape for object data},
  author={Virta, Joni},
  journal={Journal of Nonparametric Statistics},
  pages={1--21},
  year={2025},
  publisher={Taylor \& Francis}
}

@article{Stewart1980,
  author       = {G. W. Stewart},
  title        = {The Efficient Generation of Random Orthogonal Matrices with an Application to Condition Estimators},
  journal      = {SIAM Journal on Numerical Analysis},
  year         = {1980},
  volume       = {17},
  number       = {},
  pages        = {403--409}
}

@article{simeon2022riemannian,
  title={Riemannian Geometry of Functional Connectivity Matrices for Multi-Site Attention-Deficit/Hyperactivity Disorder Data Harmonization},
  author={Simeon, Guillem and Piella, Gemma and Camara, Oscar and Pareto, Deborah},
  journal={Frontiers in Neuroinformatics},
  volume={16},
  pages={769274},
  year={2022},
  publisher={Frontiers Media SA}
}

@book{bhatia2009positive,
  title={Positive Definite Matrices},
  author={Bhatia, Rajendra},
  year={2009},
  publisher={Princeton University Press}
}

@article{mahalanobis1936generalized,
  author    = {Mahalanobis, P. C.},
  title     = {On the Generalized Distance in Statistics},
  journal   = {Proceedings of the National Institute of Science (India)},
  volume    = {2},
  number    = {1},
  pages     = {49--55},
  year      = {1936}
}

@inproceedings{tukey1975mathematic,
  author    = {Tukey, John W.},
  title     = {Mathematics and the Picturing of Data},
  booktitle = {Proceedings of the International Congress of Mathematicians},
  editor    = {R. James},
  volume    = {2},
  pages     = {523--531},
  year      = {1975},
  publisher = {Canadian Mathematical Congress}
}

@article{Liu1990,
  author    = {R. Y. Liu},
  title     = {On a Notion of Data Depth Based on Random Simplices},
  journal   = {Annals of Statistics},
  volume    = {18},
  number    = {1},
  pages     = {405--414},
  year      = {1990},
}

@incollection{liu1992data,
  author    = {Liu, R. Y.},
  title     = {Data Depth and Multivariate Rank Tests},
  booktitle = {L1-Statistics Analysis and Related Methods},
  editor    = {Dodge, Y.},
  pages     = {279--294},
  year      = {1992},
  publisher = {North-Holland},
  address   = {Amsterdam}
}

@article{koshevoy1997zonoid,
  title={Zonoid Trimming for Multivariate Distributions},
  author={Koshevoy, G. and Mosler, K.},
  journal={Annals of Statistics},
  volume={25},
  number={5},
  pages={1998--2017},
  year={1997}
}

@incollection{serfling2002depth,
  author    = {Serfling, Robert},
  title     = {A Depth Function and a Scale Curve Based on Spatial Quantiles},
  booktitle = {Statistical Data Analysis Based on the L1-Norm and Related Methods},
  editor    = {Yadolah Dodge},
  series    = {Statistics for Industry and Technology},
  pages     = {25--38},
  publisher = {Birkh\"auser},
  address   = {Basel},
  year      = {2002}
}

@article{virta2023spatial,
  title={Spatial Depth for Data in Metric Spaces},
  author={Virta, Joni},
  journal={Accepted for publication in Scandinavian Journal of Statistics},
  year={2026}
}

@article{LiuModarres2011,
  author = {Liu, Z. and Modarres, R.},
  title = {Lens Data Depth and Median},
  journal = {Journal of Nonparametric Statistics},
  volume = {23},
  number = {4},
  pages = {1063--1077},
  year = {2011}
}

@article{yang2018beta,
  author = {Yang, M. and Modarres, R.},
  title = {$\beta$-skeleton Depth Functions and Medians},
  journal = {Communications in Statistics - Theory and Methods},
  volume = {47},
  number = {20},
  pages = {5127--5143},
  year = {2018},
  publisher = {Taylor \& Francis}
}

@article{dubey2024metric,
  title={ Metric Statistics: Exploration and Inference for Random Objects with Distance Profiles},
  author={Dubey, Paromita and Chen, Yaqing and M{\"u}ller, Hans-Georg},
  journal={Annals of Statistics},
  volume={52},
  number={2},
  pages={757--792},
  year={2024},
  publisher={Institute of Mathematical Statistics}
}

@article{vardi2000multivariate,
  title={ The Multivariate $L_1$-median and Associated Data Depth},
  author={Vardi, Yehuda and Zhang, Cun-Hui},
  journal={Proceedings of the National Academy of Sciences},
  volume={97},
  number={4},
  pages={1423--1426},
  year={2000},
  publisher={National Academy of Sciences}
}

@book{lee2019u,
  title={U-statistics: Theory and Practice},
  author={Lee, A J},
  year={2019},
  publisher={Routledge}
}

@book{van2000asymptotic,
  title={Asymptotic Statistics},
  author={Van der Vaart, Aad W},
  volume={3},
  year={2000},
  publisher={Cambridge University Press}
}

@article{dutta2011some,
  title={Some intriguing properties of {T}ukey’s half-space depth},
  author={Dutta, Subhajit and Ghosh, Anil K and Chaudhuri, Probal},
  journal={Bernoulli},
  volume={17},
  number={4},
  pages={1420--1434},
  year={2011}
}

@article{BFGS1995,
author = {Byrd, Richard H. and Lu, Peihuang and Nocedal, Jorge and Zhu, Ciyou},
title = {A Limited Memory Algorithm for Bound Constrained Optimization},
journal = {SIAM Journal on Scientific Computing},
volume = {16},
number = {5},
pages = {1190-1208},
year = {1995},
}

@book{burago2001metric,
  author    = {Dmitri Burago and Yuri Burago and Sergei Ivanov},
  title     = {A Course in Metric Geometry},
  year      = {2001},
  publisher = {American Mathematical Society},
  address   = {Providence, RI},
  isbn      = {0-8218-2129-6},
  series    = {Graduate Studies in Mathematics},
  volume    = {33}
}

@book{gromov2007metric,
  author    = {Mikhael Gromov},
  title     = {Metric Structures for Riemannian and Non-Riemannian Spaces},
  year      = {2007},
  publisher = {Birkhäuser},
  address   = {Boston},
  isbn      = {978-0-8176-4582-3}
}

@article{eddelbuettel2018extending,
  title={Extending {R} with {C++}: a brief introduction to {Rcpp}},
  author={Eddelbuettel, Dirk and Balamuta, James Joseph},
  journal={The American Statistician},
  volume={72},
  number={1},
  pages={28--36},
  year={2018},
  publisher={Taylor \& Francis}
}

@Manual{RMHD,
    title = {MHD: Metric Halfspace Depth},
    author = {Xiongtao Dai},
    year = {2024},
    note = {{R} package version 0.1.2},
    url = {https://CRAN.R-project.org/package=MHD},
  }

@article{mozharovskyi2025anomaly,
  title={Anomaly detection using data depth: multivariate case},
  author={Mozharovskyi, Pavlo and Valla, Romain},
  journal={International Journal of Data Science and Analytics},
  pages={1--26},
  year={2025},
  publisher={Springer}
}

@article{kleindessner2017lens,
  title={Lens depth function and $k$-relative neighborhood graph: versatile tools for ordinal data analysis},
  author={Kleindessner, Matth{\"a}us and Von Luxburg, Ulrike},
  journal={Journal of Machine Learning Research},
  volume={18},
  number={58},
  pages={1--52},
  year={2017}
}

@phdthesis{mozharovskyi:tel-03780415,
  TITLE = {{Data depth: computation, applications, and beyond}},
  AUTHOR = {Mozharovskyi, Pavlo},
  SCHOOL = {{Institut Polytechnique de Paris}},
  YEAR = {2022},
  MONTH = Jul,
  TYPE = {Habilitation thesis},
  HAL_ID = {tel-03780415},
  HAL_VERSION = {v1},
}

\end{document}